\documentclass{webofc}

\usepackage[varg]{txfonts}   % Web of Conferences font
\usepackage{hyperref}
\usepackage{url}
\hypersetup{colorlinks=true,citecolor=blue,urlcolor=blue,linkcolor=blue}
  \newcommand {\nc} {\newcommand}
  \nc {\beq} {\begin{eqnarray}}
  \nc {\eeq} {\nonumber \end{eqnarray}}
  \nc {\eeqn}[1] {\label {#1} \end{eqnarray}}
  \nc {\ve} [1] {\mbox{\boldmath $#1$}}
  \nc {\ves} [1] {\mbox{\boldmath ${\scriptstyle #1}$}}
  \nc {\mrm} [1] {\mathrm{#1}}
  \nc {\half} {\mbox{$\frac{1}{2}$}}
  \nc {\thal} {\mbox{$\frac{3}{2}$}}
  \nc {\fial} {\mbox{$\frac{5}{2}$}}
  \nc {\la} {\mbox{$\langle$}}
  \nc {\ra} {\mbox{$\rangle$}}
  \nc {\etal} {\emph{et al.}\ }
  \nc {\eq} [1] {(\ref{#1})}
  \nc {\Eq} [1] {Eq.~(\ref{#1})}
  \nc {\Sec} [1] {Sec.~\ref{#1}}
  \nc {\chap} [1] {Chapter~\ref{#1}}
  \nc {\anx} [1] {Appendix~\ref{#1}}
  \nc {\tbl} [1] {Table~\ref{#1}}
  \nc {\Fig} [1] {Fig.~\ref{#1}}
  \nc {\ex} [1] {$^{#1}$}
  \nc {\Sch} {Schr\"odinger }
  \nc {\flim} [2] {\mathop{\longrightarrow}\limits_{{#1}\rightarrow{#2}}}
  \nc {\IR} [1]{\textcolor{red}{#1}}
  \nc {\IB} [1]{\textcolor{blue}{#1}}
  \nc{\IG}[1]{\textcolor{green}{#1}}

\begin{document}
\title{Are breakup reactions sensitive to spectroscopic factors?}
\subtitle{(Spoiler alert: they ain’t)}

\author{\firstname{Live-Palm} \lastname{Kubushishi}\inst{1,2}\fnsep\thanks{\email{lkubushi@uni-mainz.de}} \and
        \firstname{Pierre} \lastname{Capel}\inst{2}\fnsep\thanks{\email{pcapel@uni-mainz.de}}
}

\institute{Institute of Nuclear and Particle Physics and Department of Physics and Astronomy, Ohio University, Athens, OH
45701,USA
\and
Institut f\"ur Kernphysik, Johannes Gutenberg-Universit\"at, 55099 Mainz, Germany}

\abstract{Halo nuclei are mostly studied through reactions.
In breakup reactions, the projectile is sent upon a target, which induces the dissociation of the loosely-bound halo from the core of the nucleus.
In most cases, a spectroscopic factor for the core-halo structure is inferred from experimental data.
In this contribution, we present a new analysis of the Coulomb breakup of $^{11}$Be on $^{208}$Pb at $69A$\,MeV measured at RIKEN.
We use an effective multi-channel particle-rotor model of the projectile, which accounts for the excitation of the $^{10}$Be core.
The calculations are in excellent agreement with the data, independently of the value of the spectroscopic factor.
This confirms that breakup cross sections are insensitive to this value and that spectroscopic factors cannot be inferred from such measurements.
In the near future, we plan to extend this idea to other reactions.
}
\maketitle
\section{Introduction}
\label{intro}
Halo nuclei are light, neutron-rich nuclei, which exhibit a much larger matter radius than their isobars \cite{Tan96}.
This exceptional size is understood as a threshold effect.
Being located close to---when not at---the neutron dripline, they exhibit a low binding energy for one or two neutrons.
Thanks to this loose binding the valence neutrons tunnel far into the classically-forbidden region and form a sort of diffuse halo around a compact core \cite{HJ87}.
This exotic structure enhances significantly the size of the nucleus.

Being short-lived, halo nuclei are mostly studied indirectly through reactions at Radioactive-Ion Beam (RIB) facilities.
In Coulomb-breakup reactions, the core-halo structure dissociates through the interaction of the projectile with a heavy---high $Z$---target \cite{NK12}.
Because it reveals the internal structure of the projectile and is dominated by the electromagnetic interaction, it is a clean experimental tool to study halo nuclei.
Information about the projectile structure is then inferred from the comparison between theoretical predictions and experiment.
The former is usually obtained using a pure single-particle description of the halo nucleus, in which the neutron is seen as bound to an inert core in its ground state \cite{BC12}.
The latter, of course, contains also all the other configurations, in which the core sits in excited states.
Because these configurations contribute much less to the breakup cross section, it is tempting to infer from the measured cross sections the probability to find the projectile in its main configuration, viz.\ the one in which the core is in its ground state \cite{NK12}.

It has however been argues that this \emph{Spectroscopic Factor} (SF) is not an observable, and, as such, cannot be measured \cite{FH02,MKF15,MBF17}.
In this work, we overlook this argument and ask ourselves: ``Were SFs observables, can we infer them from Coulomb-breakup cross sections?''

To answer this question, we consider a description of halo nuclei, in which the core can be in both its ground and first excited states \cite{KC25b}.
Varying the coupling strength between both core states, we can change at will the SF.
We then use this coupled-channel model within a description of Coulomb breakup at the first order of the perturbation theory \cite{KC26}.
This enables us to study for the first time how the SF for the main projectile configuration affects the calculated breakup cross section.
We apply this analysis to the Coulomb breakup of $^{11}$Be on $^{208}$Pb at $69A$\,MeV, which has been measured at RIKEN \cite{Fuk04}.

Once fitted to the Asymptotic Normalisation Constant (ANC) predicted \emph{ab initio} by Calci \etal \cite{CAL16}, all descriptions of $^{11}$Be lead to a very good agreement with experiment, independently of the SF \cite{KC26}.
Our results therefore confirm previous studies performed only at the single-particle approximation, viz.\ without core excitation, which have shown that breakup reactions are mainly peripheral, in the sense that they probe only the tail of the projectile wave function and not its interior \cite{CAP06,CAP07,CAP18}.
Therefore, whether observable or not, SFs should by no means be inferred from experimental Coulomb-breakup cross sections.

This study constitutes the first step of a campaign, which we plan to prolong by including higher-order effects and the projectile-target nuclear interaction.
The future developments will enable us to refine our conclusions and extend them to other reactions, such as nuclear-dominated breakup and knockout \cite{HC21}.

\section{Theoretical framework}\label{theory}

Usually the breakup of one-neutron halo nuclei is described within a three-body collision model \cite{BC12}.
The projectile is seen as a core $c$, assumed in its ground state, to which a neutron n is loosely bound.
Its structure is described by the single-particle Hamiltonian
\beq
H_0=T_r+V_{c\rm n}(\ve{r}),
\eeqn{e1}
where $\ve{r}$ is the $c$-n relative coordinate, and $V_{c\rm n}$ is an effective potential that simulates the interaction between the core and the halo neutron.
This potential is fitted to reproduce key structure observables, such as the one-neutron separation energy of the nucleus, the spin and parity of its ground state, and sometimes low-energy excited states.

The negative-energy eigenstates $\varphi_{n_rljm}$ of $H_0$ describe the $c$-n bound states in the partial wave $ljm$, where $l$ is the orbital angular momentum for the $c$-n relative motion, $j$ is the total angular momentum obtained from coupling $l$ with the spin of the neutron, and $m$ is the projection of $j$; $n_r$ is the number of nodes in their reduced radial wave function.
These radial wave functions are normed to unity and behave asymptotically as
\beq
u_{n_rlj}(r)\flim{r}{\infty} {\cal C}_{n_rlj}\ i\,\kappa_{n_rlj}r\ h^{(1)}_l (i\kappa_{n_rlj} r),
\eeqn{e1a}
where  ${\cal C}_{n_rlj}$ is the ANC of the single-particle state,
$\kappa_{n_rlj}$ is defined in relation to the ground state energy $E_{n_rlj}=-\hbar^2\kappa^2_{n_rlj}/2\mu$, and $h^{(1)}$ is a spherical Bessel function of the third kind \cite{AS70}.
The positive-energy states $\varphi_{kljm}$ describe the $c$-n continuum, i.e., the broken-up projectile.
In addition to $l$, $j$, and $m$, they are also identified by the $c$-n wavenumber $k$.

The target $T$ is described as a structureless particle, and its interaction with the projectile constituents is simulated by optical potentials $V_{cT}$ and $V_{{\rm n}T}$.
Within this framework, the study of the $P$-$T$ collision reduces to solving the three-body scattering problem \cite{BC12}
\beq
\left[T_R+H_0+V_{cT}(R_{cT})+V_{{\rm n}T}(R_{{\rm n}T})\right]\Psi(\ve{r},\ve{R})=E_t\ \Psi(\ve{r},\ve{R}),
\eeqn{e2}
where $\ve{R}$ is the coordinate of the projectile centre of mass relative to the target and $E_t$ is the total energy in the centre-of-mass frame.

Equation \eq{e2} must be solved with the condition that the projectile, initially in its ground state $\varphi_{n_{r0}l_0j_0m_0}$, is impinging on the target
\beq
\Psi^{(m_0)}(\ve{r},\ve{R})\flim{Z}{-\infty}e^{iKZ}\varphi_{n_{r0}l_0j_0m_0}(\ve{r}),
\eeqn{e3}
where the $Z$ axis is chosen along the incoming beam, and the wave number $K$ is related to the total energy $E_t$ and the energy of the projectile ground state $E_{n_{r0}l_0j_0}$: $E_t=\hbar^2K^2/2\mu_{PT}+E_{n_{r0}l_0j_0}$, with $\mu_{PT}$ the $P$-$T$ reduced mass.

\section{What do we probe in breakup reaction?}\label{probe}
\subsection{Spectroscopic Factors}
As mentioned earlier, reaction calculations are most often performed assuming that the projectile is initially in one given configuration, with its core in its ground state and the halo neutron in a well defined single-particle state with a wave function $\varphi_{n_{r0}l_0j_0m_0}$ normed to one.
However, reality is more complex, and we expect the projectile ground state to be in a superposition of various configurations, including ones in which the core is in one or more excited states.
In that case, the actual overlap wave function corresponding to the configuration, in which the core sits in its ground state, exhibits a norm lower than one.
The square of this norm is what we call the \emph{Spectroscopic Factor} (SF) ${\cal S}_{n_{r0}l_0j_0\otimes I^{\pi_c}}$, where $I^{\pi_c}$ are the spin and parity of the core.
It corresponds to the probability to find the projectile in that configuration.

This piece of structure information is often inferred from reaction data from the ratio \cite{HT03}
\beq
{\cal S}_{n_{r0}l_0j_0\otimes I^{\pi_c}}=\frac{\sigma^{\rm exp}}{\sigma^{\rm th}},
\eeqn{e4}
where $\sigma^{\rm exp}$ is the experimental cross section, and $\sigma^{\rm th}$ is the theoretical cross section obtained assuming the single-particle description of the projectile $n_{r_0}l_0j_0\otimes I^{\pi_c}$.
The idea behind \Eq{e4} is that the single-particle wave function $\varphi_{n_{r0}l_0j_0m_0}$, eigenstate of $H_0$ \eq{e1}, is a good estimate for the actual overlap wave function, but to a normalising factor.
The other configurations, viz.\ those in which the core sits in an excited state, are assumed to have a negligible impact on the reaction cross section, because they correspond to states in which the valence neutron is more deeply bound, and hence can less easily be broken up from the projectile.

Within this hypothesis, and thanks to the linearity of the three-body Schr\"odinger \Eq{e2}, the cross section inferred from the single-particle description of the projectile equals that of a full calculation assuming a multi-channel description of the projectile, but to the square of the norm of the actual overlap wave function, viz.\ ${\cal S}_{n_{r0}l_0j_0\otimes I^{\pi_c}}$.

It has been opposed to this vision that the notion of SF is erroneous, for they are not observables and, as such, cannot be measured \cite{FH02,MKF15,MBF17}.
Here, we follow a more practical approach: we assume that SFs are observables, and see if, in that case, they can be inferred from Coulomb-breakup measurements.

\subsection{Peripherality}

This question has been asked many times in the past because it has been shown that breakup reactions are peripheral, in the sense that they probe only the tail of the projectile wave function and that reaction calculations are insensitive to the halo description at short distances \cite{CAP07}.
In particular, it has been shown that breakup reactions are sensitive to the ground-state ANC ${\cal C}_{n_{r0}l_0j_0}$ and phaseshifts in the $c$-n continuum \cite{CAP06}.

The insensitivity of breakup to the short-range part of the projectile wave function has led to the idea of using a Halo Effective Field Theory (Halo-EFT) within reaction codes \cite{CAP18}.
Halo-EFT is based on the clear separation of scales, which we observe in halo nuclei between the tight and compact core, and the diffuse and extended halo \cite{BERT02}.
The single-particle Hamiltonian \eq{e1} can thus be expanded upon a small parameter, which corresponds to the ratio of the small scale (the size of the core) to the large scale (the size of the halo); see Ref.\,\cite{HAM17} for a recent review.
Because the short-range physics of the nucleus is neglected, the $c$-n interaction is reduced to a contact term and its derivatives.
For practical applications, these terms are regularised by a Gaussian.
At Next-to-Leading Order (NLO), it is parametrised as \cite{CAP18}
\beq
V_{c\rm n}(r)=V^{lj}_0 e^{-\frac{r^2}{2\sigma^2}}+V^{lj}_2 r^2 e^{-\frac{r^2}{2\sigma^2}},
\eeqn{e5}
where the Low-Energy Constants (LECs) $V^{lj}_0$ and $V^{lj}_2$ are fitted to reproduced the known structure observables of the nucleus, viz.\ its one-neutron separation energy $S_{\rm n}$ and ANC ${\cal C}_{n_{r0}l_0j_0}$ in the ground-state partial wave $l_0j_0$, and its phaseshift in the continuum  \cite{CAP18}.
The range $\sigma$ is an unfitted parameter, which corresponds to the short distances that are ignored in this description.
In practical cases, it is chosen about $\sigma=1$--2\,fm.

This idea has been successfully applied to the breakup of $^{11}$Be on both Pb and C at about $70A$\,MeV \cite{CAP18}.
For this, the LECs have been fitted to the \emph{ab initio} predictions of Calci \etal in Ref.\,\cite{CAL16} for the ground-state ANC in the $s_{1/2}$ partial wave, and the phaseshift in the $p_{3/2}$ and $p_{1/2}$ continua.
True to Halo-EFT at NLO, the higher partial waves have been described by plane wave, viz.\ assuming $V_{c\rm n}=0$.
Excellent agreement with the experimental data of Fukuda \etal \cite{Fuk04} was obtained for various breakup observables.
These results being independent of the range $\sigma$, they confirm the conclusion of Ref.\,\cite{CAP07} that the reaction is purely peripheral, and does not depend on the internal part of the projectile wave function.

However, all these results have been obtained with a single-particle wave function of norm one, and no coupling to other configurations.
To confirm the insensitivity of our calculations to SFs, we include here the core excitation within Halo-EFT \cite{KC25b} and run Coulomb-breakup calculation to check whether this conclusion changes in a coupled-channel approach \cite{KC26}.

\section{Including core excitation within Halo-EFT}

A more realistic description of $^{11}$Be should include the first $2^+$ excited state of $^{10}$Be.
In that vision, the wave function of its $\half^+$ ground-state reads
\beq
\Psi^{1/2^+}\left(^{11}{\rm Be}\right)= \varphi_{1s1/2}\otimes\phi^{0^+}\left(^{10}{\rm Be}\right) + \varphi_{0d5/2}\otimes\phi^{2_1^+}\left(^{10}{\rm Be}\right)+\varphi_{0d3/2}\otimes\phi^{2_1^+}\left(^{10}{\rm Be}\right).
\eeqn{e6a}
To include core excitation within Halo-EFT, we follow Nunes \etal and describe the projectile within a particle-rotor model \cite{NUNES96}.
The effective $c$-n Hamiltonian then reads \cite{KC26}
\beq
 H(\ve{r},\xi)= -\frac{\hbar^2}{2\mu}\Delta + V_{c\rm n}({\ve{r}},\xi) + h_{c}({\xi})
\eeqn{e6}
where $h_{c}$ is the intrinsic core Hamiltonian with eigenstates $\phi_{M_{c}}^{I_{c}^{\pi_c}}$.
The $c$-n interaction now depends on the internal coordinate of the core $\xi$; we treat this dependence at first order
\beq
    V_{c\rm n}(\ve{r},\xi)= V(r) + \beta\, \sigma\, Y_{2}^{0}(\hat{r}') \frac{d}{d\sigma}V(r),
\eeqn{e7}
where $V$ keeps the Gaussian form of \Eq{e5}.
In the usual particle-rotor model, $\beta$ corresponds to the deformation of the core \cite{NUNES96,BOHRMOTT69}.
We see it as a coupling strength between the different configurations.
The $Y_2^0$ spherical harmonics depends on the solid angle of the $c$-n coordinate $\ve{r'}$ in the core’s intrinsic reference frame.
It accounts for the core excitation to its first $2^+$ state.

To obtain the projectile wave functions, we expand them on the core eigenstates
\beq
    \Psi^{J^\pi M}(\ve{r},\xi)=\sum_{\alpha} i^l \frac{u_{\alpha}(r)}{r} [\mathcal{Y}_{l j}(\hat{r}) \otimes \phi_{M_{c}}^{I_{c}^{\pi_c}}(\xi)]^{J M},
\eeqn{e8}
where $\alpha=\{n_r,l,j,I,\pi_c\}$ and $\mathcal{Y}_{l j}$ is the spin-angular part of the $c$-n overlap wave function.
Such an expansion leads to the set of coupled equations for the radial wave functions
\beq
\left\{\frac{\hbar^2}{2\mu}\left[-\frac{d^2}{dr^2}+\frac{l(l+1)}{r^2}\right]+V_{\alpha \alpha}(r) + \epsilon_{\alpha}- E\right\} u_{\alpha}(r) =  -\sum_{\alpha'\neq \alpha}V_{\alpha \alpha'}(r)u_{\alpha'}(r),
\eeqn{e9}
where $\epsilon_\alpha$ is the energy of the core state.
They are solved within the R-Matrix method \cite{DESC10} on a Lagrange mesh \cite{BAYE15}, see Ref.\,\cite{KC25b}.

\begin{figure}[t]
\centering
\includegraphics[width=0.49\linewidth,clip]{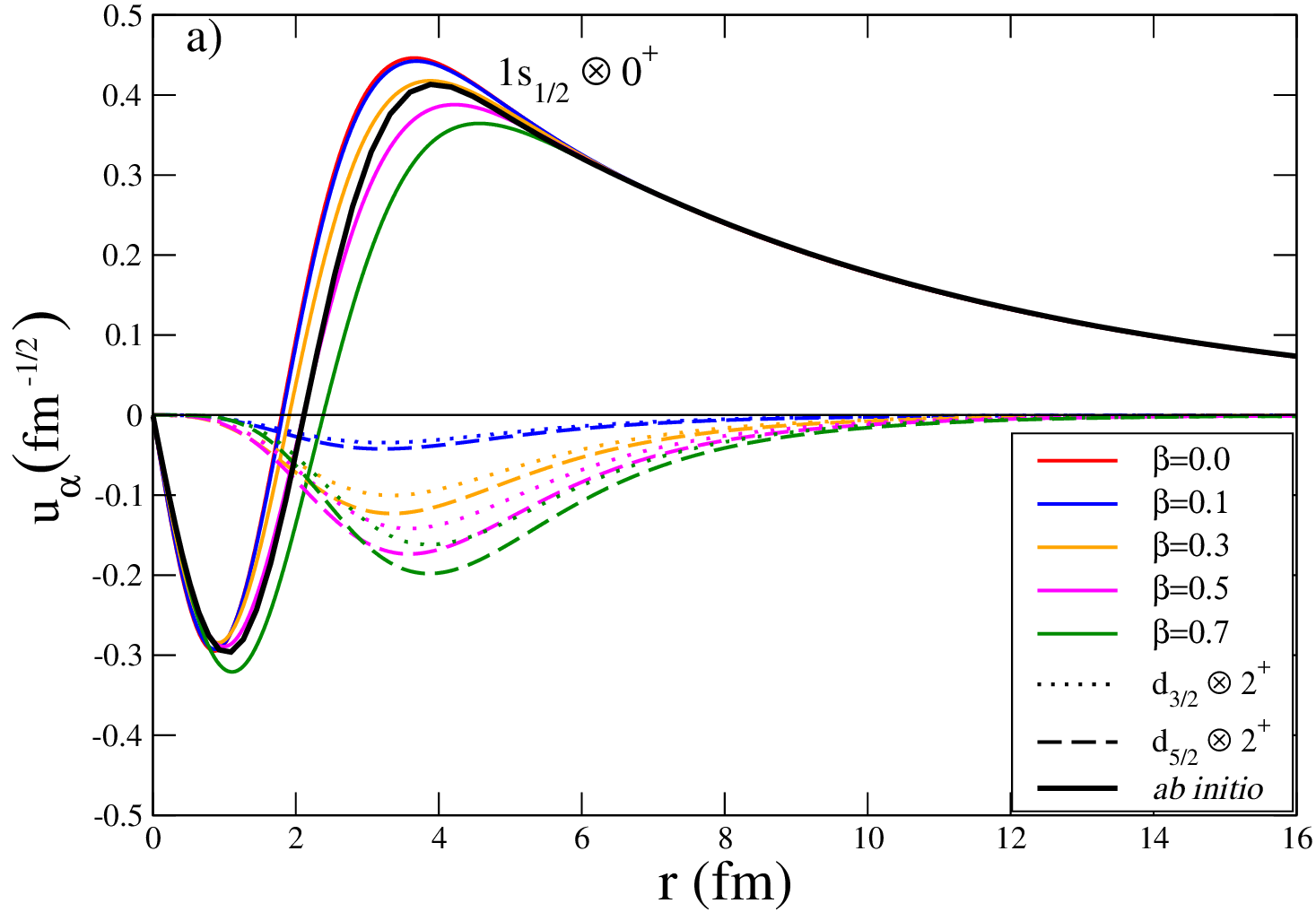}
\includegraphics[width=0.5\linewidth,clip]{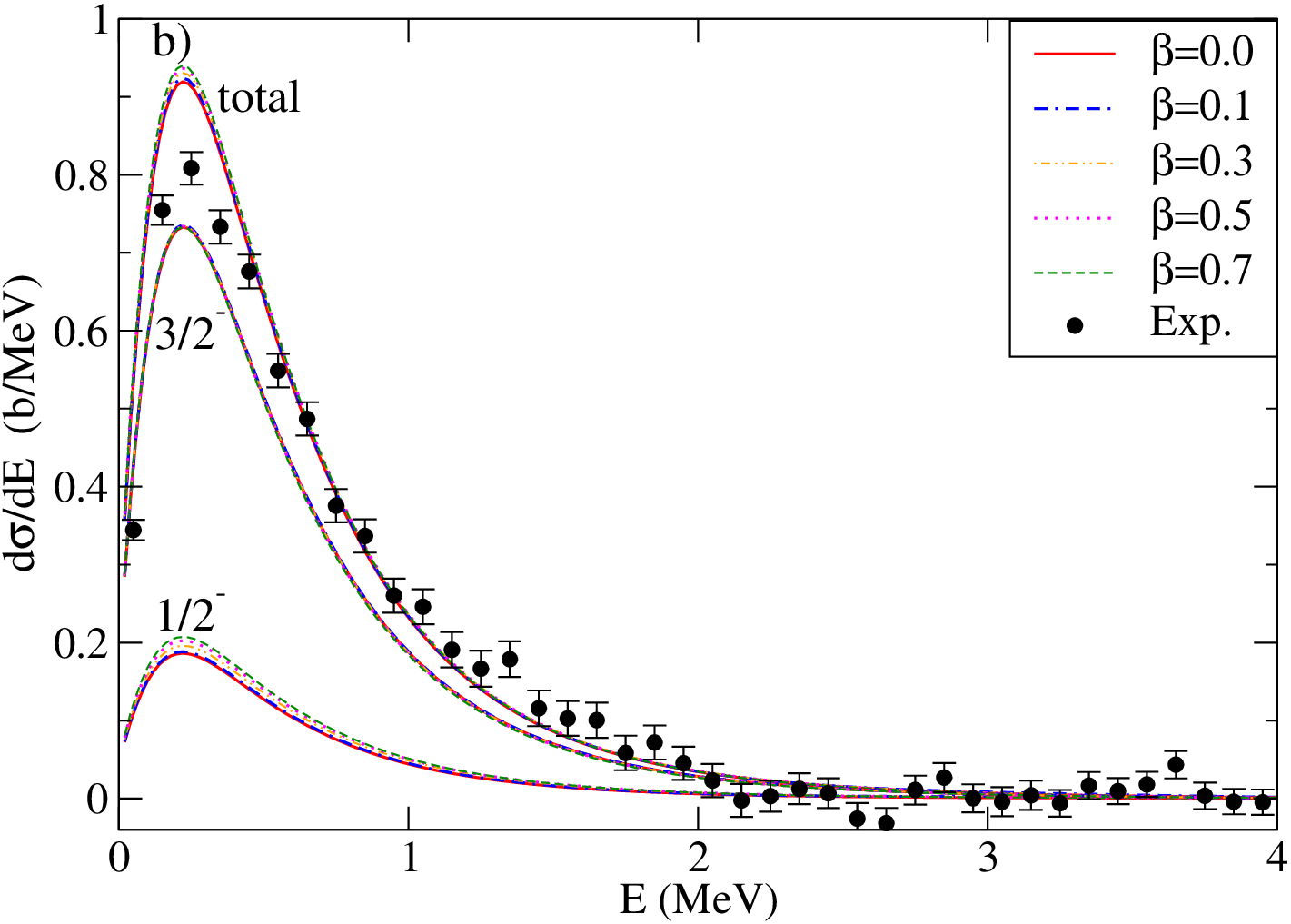}
\caption{(a) Halo-EFT description of $^{11}$Be ground state including core excitation: radial overlap wave functions obtained with different coupling strengths $\beta=0$--0.7 in the $1s_{1/2}\otimes0^+$ (solid lines), $0d_{5/2}\otimes0^+$ (dashed lines), and $0d_{3/2}\otimes0^+$ (dotted lines).
(b) Corresponding cross section for the Coulomb breakup of $^{11}$Be on $^{208}$Pb measured at $69A$\,MeV expressed as a function of the $^{10}$Be-n relative energy $E$ compared to the RIKEN data \cite{Fuk04}; the $\thal^-$ and $\half^-$ contributions are shown separately.
Figures from Ref.\,\cite{KC26}.}
\label{f1}
\end{figure}

We first follow Ref.\,\cite{CAP18} and fit the LECs of the $c$-n interaction \eq{e5} to reproduce $S_{\rm n}$ and the bound-state ANC predicted \emph{ab initio} \cite{CAL16} in the $\half^+$ partial wave, and to the \emph{ab initio} $^{10}$Be-n phaseshift in the $\thal^-$ and $\half^-$ partial waves.
We use $\sigma=2$\,fm; similar results are obtained with different values of that range \cite{KC25b}.
The radial overlap wave functions for the three configurations of \Eq{e6a} are shown in \Fig{f1}(a) for coupling strengths $\beta$ between 0, when there is no coupling (red line), and 0.7 (green lines).
Increasing $\beta$, we transfer probability strength from the dominant $1s_{1/2}\otimes 0^+$ configuration (solid lines) to the $0d_{5/2}\otimes 2^+$ (dashed lines) and $0d_{3/2}\otimes 2^+$ (dotted lines) ones.
${\cal S}_{1s_{1/2}\otimes0^+}$ decreases from 1 at $\beta=0$ to 0.8 at $\beta=0.7$.

We then use this description of the projectile's ground state and the corresponding continuum \cite{KC25b} to compute the Coulomb-breakup cross section of $^{11}$Be on $^{208}$Pb at $69A$\,MeV \cite{KC26}.
In this first analysis within a coupled-channel model of one-neutron halo nuclei, we describe the reaction within a simple first-order of the perturbation theory, viz.\ Alder and Winther \cite{ALDWIN75}.
The results are shown in \Fig{f1}(b) in comparison with the experimental data of Fukuda \etal selected at forward angle \cite{Fuk04}.

Despite the 20\% difference in SF, there is no change in the calculation of the breakup cross section for $^{11}$Be on $^{208}$Pb at $69A$\,MeV shown in \Fig{f1}(b).
This confirms, for the first time within a coupled-channel description of the projectile, that Coulomb breakup is purely peripheral, and that it does not depend upon the norm of the overlap wave function of the halo configuration, viz.\ the so-called SF.
It also confirms the value predicted \emph{ab initio} for the ANC ${\cal C}_{1s_{1/2}\otimes0^+}$, since it leads to good agreement with the RIKEN data of Ref.\,\cite{Fuk04}.

\begin{figure}[t]
\centering
\includegraphics[width=0.49\linewidth,clip]{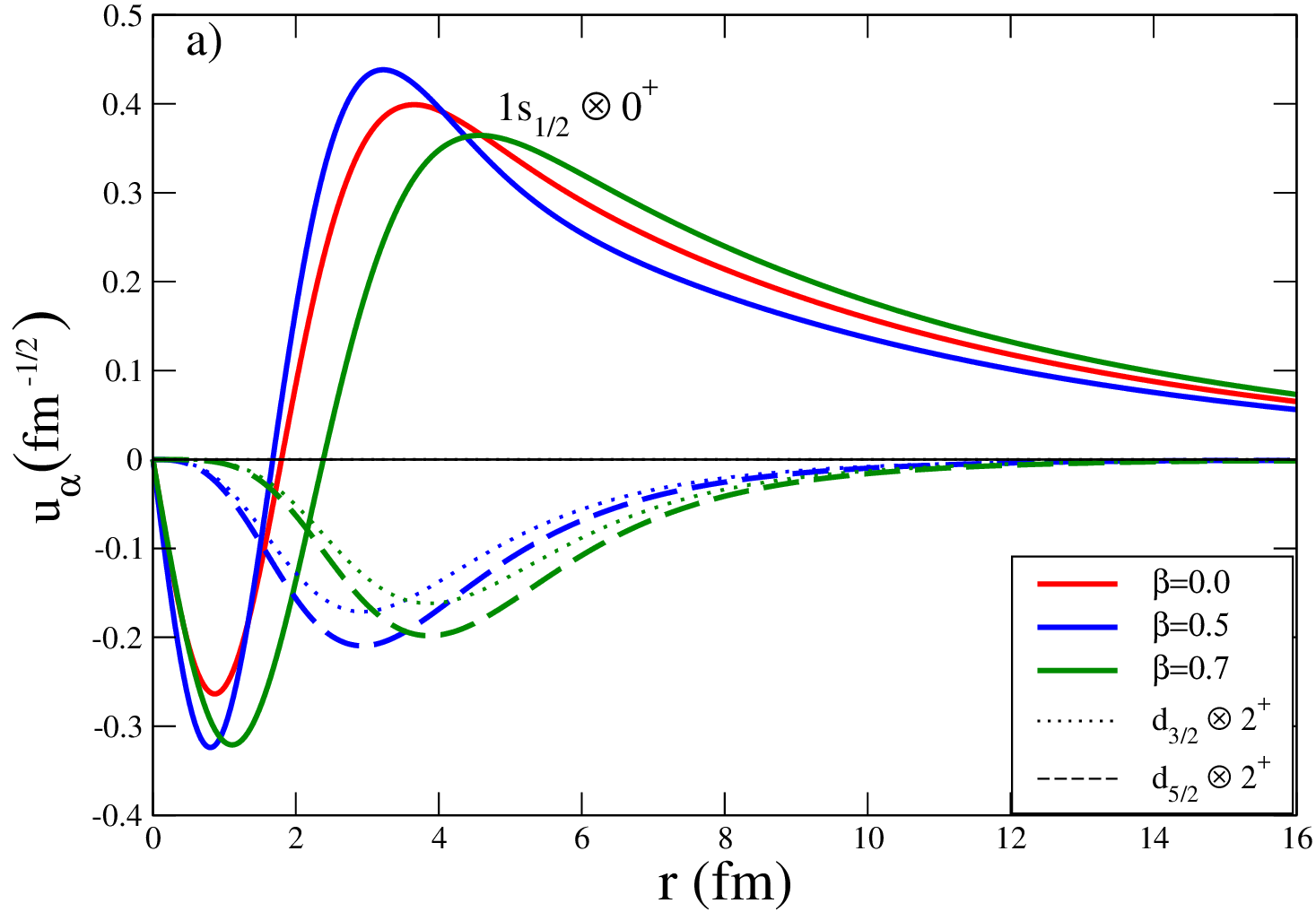}
\includegraphics[width=0.49\linewidth,clip]{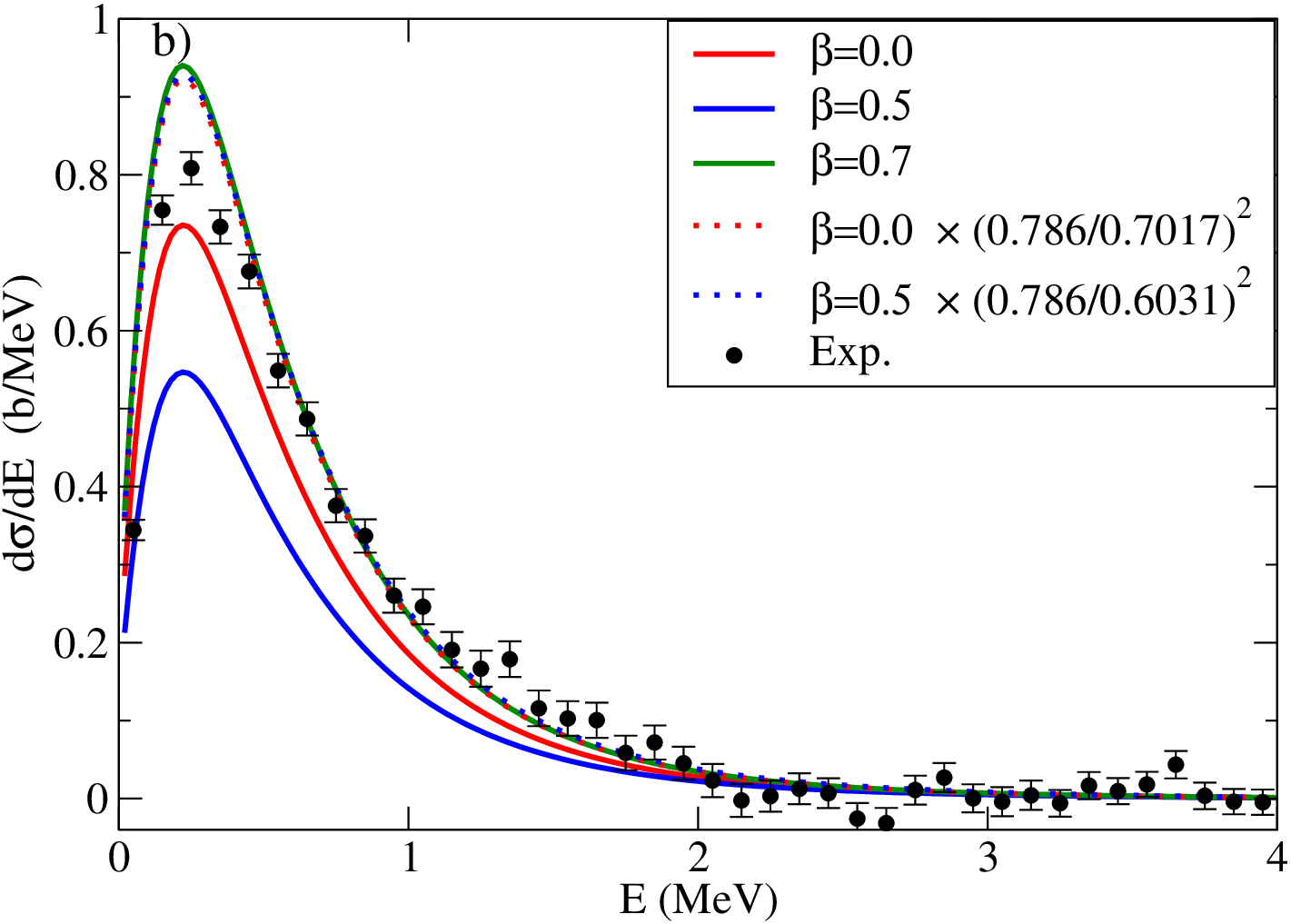}
\caption{Insensitivity of the breakup cross section to the SF for $^{11}$Be on $^{208}$Pb at $69A$\,MeV.
(a) Radial overlap wave function of the Halo-EFT description of $^{11}$Be with core excitation fixing the LECs to ${\cal S}_{1s_{1/2}\otimes0^+}=0.8$ for $\beta=0$ (red), 0.5 (blue), and  0.7 (green).
(b) The corresponding Coulomb breakup cross sections (solid lines) differ significantly from each other.
Good agreement with the data from Ref.\,\cite{Fuk04} is observed only for $\beta=0.7$, which reproduces the \emph{ab initio} ANC of Ref.\,\cite{CAL16}.
When scaled to that ANC, the results for $\beta=0$ and 0.5 (dotted lines) also reproduce the data.
 Figures from Ref.\,\cite{KC26}.}
\label{f3}
\end{figure}

We then perform another series of calculations, in which we fit the LECs to reproduce no longer the \emph{ab initio} ANC, but to fit ${\cal S}_{1s_{1/2}\otimes0^+}=0.8$.
The corresponding radial overlap wave functions are shown in \Fig{f3}(a) for $\beta=0$ (red line), 0.5 (blue lines), and 0.7 (blue lines).
In this test, the ANC varies: ${\cal C}_{1s_{1/2}\otimes0^+}=0.702$\,fm$^{-1/2}$ for $\beta=0$, 0.603\,fm$^{-1/2}$ for $\beta=0.5$, and 0.786\,fm$^{-1/2}$ for $\beta=0.7$, which corresponds to the \emph{ab initio} prediction of Ref.\,\cite{CAL16}.

Although all three descriptions are fitted to produce the same SF for the main $1s_{1/2}\otimes0^+$ configuration, they do not produce the same Coulomb-breakup cross section.
In \Fig{f3}(b), we observe a difference in the cross section up to 40\%.
This difference is perfectly explained by the changes observed in the square of the ANC $|{\cal C}_{1s_{1/2}\otimes0^+}|^2$.
Once scaled to the \emph{ab initio} ANC ${\cal S}_{1s_{1/2}\otimes0^+}=0.786$\,fm$^{-1/2}$, the $\beta=0$ (red dotted line) and $\beta=0.5$ (blue dotted line) are superimposed on the calculation with $\beta=0.7$ fitted to the ANC prediction of Ref.\,\cite{CAL16} (green solid line).
With that value of the ANC, they are in good agreement with the data, as already observed in one-channel calculations \cite{CAP18}.

\section{Conclusion}
Coulomb breakup is a usual experimental tool to study the structure of halo nuclei \cite{NK12}.
Often a spectroscopic factor is inferred from the comparison of measurements and theoretical predictions obtained from a single-particle description of the projectile.

However, it has been shown that breakup reactions are mostly peripheral, in the sense that they probe only the tail of the projectile wave function \cite{CAP07}.
Accordingly, they are sensitive to the ANC of the ground-state wave function and to the phaseshift in the $c$-n continuum \cite{CAP06}, and therefore potentially insensitive to the norm of the core-neutron overlap wave function.
Nevertheless, these former studies have been performed with models of reactions, in which the projectile is always described within a single-particle approach with a wave function normed to unity.
In this presentation, we have shown the first analysis of the insensitivity of Coulomb-breakup cross sections to SFs using a coupled-channel description of the projectile.
That description is based on a Halo-EFT similar to the one used in Ref.\,\cite{CAP18}, in which we take into account the excitation of the core to its first excited state \cite{KC25b}.

We apply this analysis to the case of the Coulomb breakup of $^{11}$Be on $^{208}$Pb, which has been measured at RIKEN at $69A$\,MeV \cite{Fuk04}.
In this first step, we describe this reaction at the first order of the perturbation theory \cite{KC26}.

We confirm the previous studies that this reaction is purely peripheral \cite{CAP07,CAP18}.
However, doing so in a coupled-channel method, we have, for the first time, shown that the breakup cross section is insensitive to the value of the SF.
Once the ANC of the ground-state wave function has been fixed, the outcome of the reaction calculation is independent of the norm of the main $1s_{1/2}\otimes0^+$ channel.
In addition, we have shown that wave functions with the same SF but different ANCs lead to different breakup cross sections; those cross sections scale perfectly to $|{\cal C}_{1s_{1/2}\otimes0^+}|^2$.

This confirms that the main structure observable probed in Coulomb-breakup reactions is the ANC of the ground-state wave function and not its SF.
Accordingly, SF should by no means be inferred from such reactions.
Rather these experimental cross sections can be used to test ANCs predicted from accurate structure calculations \cite{CAL16}.
Note that the value of the ANC hence inferred depends also on the description of the continuum, and more particularly on the phaseshift in the dominant partial waves reached through the Coulomb excitation of the ground state \cite{TYPBAU05,CAP06,CAP18}.

In the future, we will extend this endeavour making use of more sophisticated reaction models to go beyond the first order of the perturbation theory.
This will enable us to properly account for higher-order effects, such as coupling within the continuum \cite{CB05}, and include the nuclear interactions between the projectile constituents and the target.
This latter step will enable us in particular to extend this study to nuclear-dominated reactions, such as breakup or knockout on light targets \cite{HC21}.

\section*{Acknowledgments}
This work was supported by the Deutsche Forschungsgemeinschaft (DFG, German Research Foundation) through Project-ID 279384907 – SFB 1245 and the Cluster of Excellence “Precision Physics, Fundamental Interactions and Structure of Matter” (PRISMA++ EXC 2118/2, Project ID No. 390831469), and by the U.S.~Department of Energy under contract No.~DE-FG02-93ER40756.


\begin{thebibliography}{26}

\bibitem{Tan96}
I.~Tanihata, Neutron halo nuclei, J. Phys. G \textbf{22}, 157 (1996).
  \doiwoc{10.1088/0954-3899/22/2/004}

\bibitem{HJ87}
P.~G.~Hansen, B.~Jonson, The neutron halo of extremely neutron-rich nuclei,
  Europhys. Lett. \textbf{4}, 409 (1987). \doiwoc{10.1209/0295-5075/4/4/005}

\bibitem{NK12}
T.~Nakamura, Y.~Kondo, Neutron halo and breakup reactions, Lect. Not. Phys.
  \textbf{848}, 67 (2012), {Ed. C. Beck}. \doiwoc{10.1007/978-3-642-24707-1_2}

\bibitem{BC12}
D.~Baye, P.~Capel, Breakup reaction models for two- and three-cluster
  projectiles, Lect. Not. Phys. \textbf{848}, 121 (2012), {Ed. C. Beck}.
  \doiwoc{10.1007/978-3-642-24707-1_3}

\bibitem{FH02}
R.~Furnstahl, H.-W.~Hammer, Are occupation numbers observable?, Phys. Lett. B
  \textbf{531}, 203 (2002).
  \doiwoc{https://doi.org/10.1016/S0370-2693(01)01504-0}

\bibitem{MKF15}
S.~N.~More, S.~K\"onig, R.~J.~Furnstahl, K.~Hebeler, Deuteron
  electrodisintegration with unitarily evolved potentials, Phys. Rev. C
  \textbf{92}, 064002 (2015). \doiwoc{10.1103/PhysRevC.92.064002}

\bibitem{MBF17}
S.~N.~More, S.~K.~Bogner, R.~J.~Furnstahl, Scale dependence of deuteron
  electrodisintegration, Phys. Rev. C \textbf{96}, 054004 (2017).
  \doiwoc{10.1103/PhysRevC.96.054004}

\bibitem{KC25b}
L.-P.~Kubushishi, P.~Capel, Exploring core excitation in halo nuclei using halo
  effective field theory: an application to the bound states of {$^{11}$Be}
  (2025), \texttt{2507.13585},
  \urlstyle{tt}\url{https://arxiv.org/abs/2507.13585}

\bibitem{KC26}
L.-P.~Kubushishi, P.~Capel, Insensitivity of the {Coulomb} breakup of halo
  nuclei to spectroscopic factors, Phys. Lett. B \textbf{879}, 140694 (2026).
  \doiwoc{https://doi.org/10.1016/j.physletb.2026.140694}

\bibitem{Fuk04}
N.~Fukuda, T.~Nakamura, N.~Aoi, N.~Imai, M.~Ishihara, T.~Kobayashi, H.~Iwasaki,
  T.~Kubo, A.~Mengoni, M.~Notani et~al., Coulomb and nuclear breakup of a halo
  nucleus $^{11}\mathrm{Be}$, Phys. Rev. C \textbf{70}, 054606 (2004).
  \doiwoc{10.1103/PhysRevC.70.054606}

\bibitem{CAL16}
A.~Calci, P.~Navr\'atil, R.~Roth, J.~Dohet-Eraly, S.~Quaglioni, G.~Hupin, Can
  ab initio theory explain the phenomenon of parity inversion in
  $^{11}\mathrm{Be}$?, Phys. Rev. Lett. \textbf{117}, 242501 (2016).
  \doiwoc{10.1103/PhysRevLett.117.242501}

\bibitem{CAP06}
P.~Capel, F.M. Nunes, Influence of the projectile description on breakup
  calculations, Phys. Rev. C \textbf{73}, 014615 (2006).
  \doiwoc{10.1103/PhysRevC.73.014615}

\bibitem{CAP07}
P.~Capel, F.M. Nunes, Peripherality of breakup reactions, Phys. Rev. C
  \textbf{75}, 054609 (2007). \doiwoc{10.1103/PhysRevC.75.054609}

\bibitem{CAP18}
P.~Capel, D.~R.~Phillips, H.-W.~Hammer, Dissecting reaction calculations using
  halo effective field theory and ab initio input, Phys. Rev. C \textbf{98},
  034610 (2018). \doiwoc{10.1103/PhysRevC.98.034610}

\bibitem{HC21}
C.~Hebborn, P.~Capel, Halo effective field theory analysis of one-neutron
  knockout reactions of $^{11}\mathrm{Be}$ and $^{15}\mathrm{C}$, Phys. Rev. C
  \textbf{104}, 024616 (2021). \doiwoc{10.1103/PhysRevC.104.024616}

\bibitem{AS70}
M.~Abramowitz, I.~A.~Stegun, Handbook of Mathematical Functions (Dover,
  New-York, 1970)

\bibitem{HT03}
P.~G.~Hansen, J.~A.~Tostevin, Direct reactions with exotic nuclei, Annu. Rev.
  Nucl. Part. Sci. \textbf{53}, 219 (2003).

\bibitem{BERT02}
C.~Bertulani, H.-W.~Hammer, U.~{van Kolck}, Effective field theory for halo
  nuclei: shallow p-wave states, Nuclear Physics A \textbf{712}, 37 (2002).
  \doiwoc{https://doi.org/10.1016/S0375-9474(02)01270-8}

\bibitem{HAM17}
H.-W. Hammer, C.~Ji, D.~R.~Phillips, Effective field theory description of halo
  nuclei, J. Phys. G \textbf{44}, 103002 (2017).
  \doiwoc{10.1088/1361-6471/aa83db}

\bibitem{NUNES96}
F.~Nunes, I.~Thompson, R.~Johnson, Core excitation in one neutron halo systems,
  Nucl. Phys. A \textbf{596}, 171 (1996).
  \doiwoc{https://doi.org/10.1016/0375-9474(95)00398-3}

\bibitem{BOHRMOTT69}
A.~Bohr, B.~Mottelson, Nuclear Structure (W. A. Benjamin, New York, 1969)

\bibitem{DESC10}
P.~Descouvemont, D.~Baye, {The R-matrix theory}, Rep. Prog. Phys. \textbf{73},
  036301 (2010). \doiwoc{10.1088/0034-4885/73/3/036301}

\bibitem{BAYE15}
D.~Baye, {The Lagrange-mesh method}, Phys. Rep. \textbf{565}, 1 (2015).
  \doiwoc{https://doi.org/10.1016/j.physrep.2014.11.006}

\bibitem{ALDWIN75}
K.~Alder, A.~Winther, Electromagnetic Excitation: Theory of Coulomb Excitation
  with Heavy Ions (North-Holland Publishing Company, 1975)

\bibitem{TYPBAU05}
S.~Typel, G.~Baur, Electromagnetic strength of neutron and proton
  single-particle halo nuclei, Nucl. Phys. A \textbf{759}, 247 (2005).
  \doiwoc{https://doi.org/10.1016/j.nuclphysa.2005.05.145}

\bibitem{CB05}
P.~Capel, D.~Baye, {Coupling-in-the-continuum effects in Coulomb dissociation
  of halo nuclei}, Phys. Rev. C \textbf{71}, 044609 (2005).
  \doiwoc{10.1103/PhysRevC.71.044609}

\end{thebibliography}
\end{document}